\documentclass[conference,twocolumn,10pt]{IEEEtran}
\ifCLASSINFOpdf
  \usepackage[pdftex]{graphicx}
\else
\fi
\usepackage{amsmath}
  \usepackage[caption=false,font=footnotesize]{subfig}
\usepackage{url}
\usepackage[linesnumbered,ruled]{algorithm2e}

\newcommand{\comment}[1]{{}}

\renewcommand{\j}{\textrm{j}}

\usepackage{amssymb}
\usepackage{xspace}

\newcommand{\complex}{\ensuremath{\mathbb{C}}\xspace}
\newcommand{\complexpow}[2]{\ensuremath{\complex^{#1 \times #2}}\xspace}

\renewcommand{\j}{\ensuremath{\textrm{j}}}

\usepackage{bm}

\newcommand{\inv}{\ensuremath{^{-1}}\xspace}

\newcommand{\ctrans}{\ensuremath{^{{*}}}\xspace}

\newcommand{\mat}[1]{\ensuremath{\mathbf{#1}}\xspace} % matrices
\renewcommand{\vec}[1]{\ensuremath{\mathbf{#1}}\xspace} % vectors
\newcommand{\trace}{\ensuremath{\textrm{tr}}}

\newcommand{\entry}[2]{\ensuremath{\left[{#1}\right]_{#2}}\xspace}

\newcommand{\cgauss}[2]{\ensuremath{\mathcal{N}_{\complex} \left( {#1} , {#2} \right) }\xspace} % complex gaussian/normal distribution
\newcommand{\ev}[1]{\ensuremath{\mathbb{E}\left[{#1}\right]}\xspace}

\newcommand{\pre}{\ensuremath{\mat{F}}\xspace} 				   % precoder
\newcommand{\prebb}{\ensuremath{\mat{F}_{\textrm{BB}}}\xspace} % baseband precoder
\newcommand{\prerf}{\ensuremath{\mat{F}_{\textrm{RF}}}\xspace} % RF precoder
\newcommand{\combb}{\ensuremath{\mat{W}_{\textrm{BB}}}\xspace} % baseband combiner
\newcommand{\comrf}{\ensuremath{\mat{W}_{\textrm{RF}}}\xspace} % RF combiner

\newcommand{\channel}{\ensuremath{\mat{H}}\xspace} % MIMO channel
\newcommand{\channeleff}{\ensuremath{\tilde{\mat{H}}}\xspace} % MIMO channel

\newcommand{\Nt}{\ensuremath{N_{\mathrm{t}}}\xspace} % number of antennas at Tx
\newcommand{\Nr}{\ensuremath{N_{\mathrm{r}}}\xspace} % number of antennas at Rx
\newcommand{\Np}{\ensuremath{N_{\mathrm{p}}}\xspace} % number of point targets

\newcommand{\Nrftx}{\ensuremath{L_{\textrm{t}}}\xspace} % number of RF chains at Tx
\newcommand{\Nrfrx}{\ensuremath{L_{\textrm{r}}}\xspace} % number of RF chains at Rx

\newcommand{\Ns}{\ensuremath{N_{\textrm{s}}}\xspace} % number of spatial streams
\newcommand{\Pt}{\ensuremath{P_{\textrm{tx}}\xspace}}
\newcommand{\Rs}{\ensuremath{\mat{R}_{\textrm{s}}\xspace}}
\newcommand{\Rn}{\ensuremath{\mat{R}_{\textrm{n}}\xspace}}

\newcommand{\symbvec}{\ensuremath{\vec{s}}\xspace}
\newcommand{\noisevec}{\ensuremath{\vec{n}}\xspace}
\newcommand{\rxvec}{\ensuremath{\hat{\vec{s}}}\xspace}

\newcommand{\numrays}{\ensuremath{N_{\mathrm{rays}}}\xspace}
\newcommand{\numclust}{\ensuremath{N_{\mathrm{clust}}}\xspace}

\newcommand{\frobenius}[1]{\ensuremath{\left\lVert#1\right\rVert_{\textsf{F}}}}
\newcommand{\frobeniustwo}[1]{\ensuremath{\left\lVert#1\right\rVert^{2}_{\textsf{F}}}}

\newcommand{\aoa}{\ensuremath{\theta}\xspace}
\newcommand{\aod}{\ensuremath{\phi}\xspace}

\newcommand{\rxresponsevector}{\ensuremath{\vec{a}_{\textrm{r}}(\azimuth_{i,j},\elevation_{i,j})}\xspace}
\newcommand{\txresponsevector}{\ensuremath{\vec{a}_{\textrm{t}}(\azimuth_{i,j},\elevation_{i,j})}\xspace}

\renewcommand{\rxresponsevector}{\ensuremath{\vec{a}_{\textrm{r}}}\xspace}
\renewcommand{\txresponsevector}{\ensuremath{\vec{a}_{\textrm{t}}}\xspace}

\newcommand{\rxresponsevectorradar}{\ensuremath{\vec{a}_{\mathrm{r,rad}}}\xspace}
\newcommand{\txresponsevectorradar}{\ensuremath{\vec{a}_{\mathrm{t,rad}}}\xspace}

\newcommand{\snr}{\ensuremath{\mathrm{SNR}}\xspace}

\newcommand{\sir}{\ensuremath{\mathrm{SIR}}\xspace}

\newcommand{\node}[1]{\unskip\ensuremath{^{^{\left(#1\right)}}}\xspace}

\newcommand{\ctransnode}[1]{\ensuremath{^{{*}^{\left(#1\right)}}}\xspace}

\usepackage{xspace}
\usepackage[acronym,nogroupskip,nonumberlist,nopostdot]{glossaries}
\makeglossaries

\usepackage{xcolor}

\newacronym{snr}{SNR}{signal-to-noise ratio}
\newacronym{sinr}{SINR}{signal-to-interference-plus-noise ratio}
\newacronym{inr}{INR}{interference-to-noise ratio}
\newacronym{sir}{SIR}{signal-to-interference ratio}
\newacronym{ian}{IAN}{interference as noise}
\newacronym{ber}{BER}{bit error rate}
\newacronym{pn}{PN}{pseudorandom noise}
\newacronym{bfsk}{BFSK}{binary frequency shift keying}
\newacronym{fh}{FH}{frequency-hopped}
\newacronym{fh-bfsk}{FH-BFSK}{frequency-hopped binary frequency shift keying}
\newacronym{crc}{CRC}{cyclic redundancy check}
\newacronym{isi}{ISI}{intersymbol interference}
\newacronym{dsss}{DSSS}{direct-sequence spread spectrum}
\newacronym{ofdm}{OFDM}{orthogonal frequency-division multiplexing}
\newacronym{ofdma}{OFDMA}{orthogonal frequency-division multiple access}
\newacronym{sdr}{SDR}{software-defined radio}
\newacronym{tx}{TX}{transmitter}
\newacronym{rx}{RX}{receiver}
\newacronym{fdd}{FDD}{frequency-division duplexing}
\newacronym{tdd}{TDD}{time-division duplexing}
\newacronym{fdma}{FDMA}{frequency-division multiple access}
\newacronym{tdma}{TDMA}{time-division multiple access}
\newacronym{sdma}{SDMA}{space-division multiple access}
\newacronym[plural=MPCs]{mpc}{MPC}{multipath component}
\newacronym{mui}{MUI}{multi-user interference}

\newacronym{fmcw}{FMCW}{frequency-modulated continuous-wave}
\newacronym{ti}{TI}{Texas Instruments}
\newcommand{\ti}{\gls{ti}\xspace}
\newacronym{cdf}{CDF}{cumulative density function}

\newacronym{rzf}{RZF}{regularized zero forcing}
\newacronym{ls}{LS}{least-squares}
\newacronym{lms}{LMS}{least mean squares}
\newacronym{rls}{RLS}{recursive least-squares}
\newacronym{mmse}{MMSE}{minimum mean square error}
\newacronym{lmmse}{LMMSE}{linear \acrlong{mmse}}
\newacronym{mse}{MSE}{mean square error}
\newacronym{fft}{FFT}{fast Fourier transform}
\newacronym{dft}{DFT}{discrete Fourier transform}
\newacronym{dtft}{DTFT}{discrete-time Fourier transform}
\newacronym{ctft}{CTFT}{continuous-time Fourier transform}
\newacronym{ml}{ML}{machine learning}
\newacronym[plural=NNs]{nn}{NN}{neural network}
\newacronym[plural=RNNs]{rnn}{RNN}{recurrent neural network}
\newacronym[plural=ADCs]{adc}{ADC}{analog-to-digital converter}
\newacronym[plural=DACs]{dac}{DAC}{digital-to-analog converter}
\newacronym[plural=FPGAs]{fpga}{FPGA}{field-programmable gate array}
\newacronym{evm}{EVM}{error vector magnitude}
\newacronym{psd}{PSD}{power spectral density}
\newacronym{enob}{ENOB}{effective number of bits}
\newacronym{zf}{ZF}{zero-forcing}
\newacronym{rv}{r.v.}{random variable}
\newacronym{omp}{OMP}{orthogonal matching pursuit}
\newacronym{svd}{SVD}{singular value decomposition}

\newacronym{agc}{AGC}{automatic gain control}
\newacronym{rf}{RF}{radio frequency}
\newacronym{los}{LOS}{line-of-sight}
\newacronym{nlos}{NLOS}{non-line-of-sight}
\newacronym{ple}{PLE}{path loss exponent}
\newacronym[plural=dB]{db}{dB}{decibel}
\newacronym{pa}{PA}{power amplifier}
\newacronym{lna}{LNA}{low noise amplifier}
\newacronym{cw}{CW}{continuous wave}
\newacronym{papr}{PAPR}{peak-to-average power ratio}
\newacronym{usrp}{USRP}{Universal Software Radio Peripheral}
\newacronym{irr}{IRR}{image rejection ratio}
\newacronym{lo}{LO}{local oscillator}
\newacronym{vm}{VM}{vector modulator}
\newacronym{mmwave}{mmWave}{millimeter-wave}

\newacronym{ism}{ISM}{industrial, scientific, and medical}
\newacronym{csma}{CSMA}{carrier-sense multiple access}
\newacronym{csmaca}{CSMA/CA}{carrier-sense multiple access with collision avoidance}
\newacronym{csmacd}{CSMA/CD}{carrier-sense multiple access with collision detection}
\newacronym{mac}{MAC}{medium access control}
\newacronym{phy}{PHY}{physical layer}
\newacronym{4g}{4G}{fourth generation}
\newacronym{lte}{LTE}{Long-Term Evolution}
\newacronym{4glte}{4G LTE}{\gls{4g} \gls{lte}}
\newacronym{5g}{5G}{fifth generation}
\newacronym{nr}{NR}{New Radio}
\newacronym{5gnr}{5G NR}{\gls{5g} \gls{nr}}
\newacronym{ieee}{IEEE}{Institute of Electrical and Electronics Engineers}
\newacronym{wifi}{Wi-Fi}{IEEE 802.11}
\newacronym{lan}{LAN}{local area network}
\newacronym{wlan}{WLAN}{wireless local area network}
\newacronym[plural=BSs]{bs}{BS}{base station}
\newacronym[plural=SBSs]{sbs}{SBS}{small-cell base station}
\newacronym[plural=FD-SBSs]{fdsbs}{FD-SBS}{\gls{fd}-enabled \gls{sbs}}
\newacronym[plural=MBSs]{mbs}{MBS}{macrocell base station}
\newacronym[plural=UEs]{ue}{UE}{user equipment}
\newacronym{ul}{UL}{uplink}
\newacronym{dl}{DL}{downlink}
\newacronym{qos}{QoS}{Quality of Service}
\newacronym{fcc}{FCC}{Federal Communications Commission}
\newacronym{iab}{IAB}{integrated access and backhaul}
\newacronym{fab}{FAB}{fixed access and backhaul}
\newacronym{hetnet}{HetNet}{heterogeneous network}

\newacronym{rcs}{RCS}{radar cross section}

\newacronym{siso}{SISO}{single-input single-output}
\newacronym{mimo}{MIMO}{multiple-input multiple-output}
\newacronym{sumimo}{SU-MIMO}{single-user \gls{mimo}}
\newacronym{mumimo}{MU-MIMO}{multi-user \gls{mimo}}
\newacronym{bf}{BF}{beamforming}
\newacronym{ca}{CA}{constant amplitude}
\newacronym{ula}{ULA}{uniform linear array}
\newacronym{aoa}{AoA}{angle of arrival}
\newacronym{aod}{AoD}{angle of departure}
\newacronym{dof}{DoF}{degrees of freedom}
\newacronym{csi}{CSI}{channel state information}
\newacronym{csit}{CSIT}{\gls{csi} at the transmitter}
\newacronym{csir}{CSIR}{\gls{csi} at the receiver}

\newacronym{fd}{FD}{in-band full-duplex}
\newacronym{hd}{HD}{half-duplex}
\newacronym{si}{SI}{self-interference}
\newacronym{sic}{SIC}{self-interference cancellation}
\newacronym{soi}{SoI}{signal of interest}
\newacronym{asic}{A-SIC}{analog \gls{sic}}
\newacronym{dsic}{D-SIC}{digital \gls{sic}}
\newacronym{star}{STAR}{simultaneous transmit and receive}
\newacronym{warp}{WARP}{Wireless Open-Access Research Platform}
\newacronym{bfc}{BFC}{beamforming cancellation}
\newacronym{ipi}{IPI}{inter-panel-interference}
\newacronym{ipic}{IPIC}{inter-panel-interference cancellation}

\newacronym{qcqp}{QCQP}{quadratically-constrained quadratic programming}

\newacronym{elf}{ELF}{extremely low frequency}
\newacronym{slf}{SLF}{super low frequency}
\newacronym{ulf}{ULF}{ultra low frequency}
\newacronym{vlf}{VLF}{very low frequency}
\newacronym{lf}{LF}{low frequency}
\newacronym{mf}{MF}{medium frequency}
\newacronym{hf}{HF}{high frequency}
\newacronym{vhf}{VHF}{very high frequency}
\newacronym{uhf}{UHF}{ultra high frequency}
\newacronym{shf}{SHF}{super high frequency}
\newacronym{ehf}{EHF}{extremely high frequency}
\newacronym{thf}{THF}{tremendously high frequency}

\newacronym{wncg}{WNCG}{Wireless Networking and Communications Group}
\newacronym{linc}{LINC}{Laboratory of Informatics, Networks, and Communications}
\newacronym{ut}{UT Austin}{The University of Texas at Austin}
\newacronym{uiuc}{UIUC}{University of Illinois at Urbana-Champaign}
\newacronym{usc}{USC}{University of Southern California}
\newacronym{mit}{MIT}{Massachusetts Institute of Technology}
\newacronym{berkeley}{UC Berkeley}{University of California, Berkeley}
\newacronym{osu}{OSU}{Ohio State University}

\newcommand{\mmwave}{\gls{mmwave}\xspace}

\newcommand{\mimo}{\gls{mimo}\xspace}

\newcommand{\rf}{\gls{rf}\xspace}

\newcommand{\fg}{\gls{5g}\xspace}

\newcommand{\figref}[1]{\figurename~\ref{#1}}

\newcounter{mytempeqncnt}

\begin{document}

% \title{Title\\Here}
\title{Enabling In-Band Coexistence of Millimeter-Wave \\ Communication and Radar}

\author{Hardik B. Jain, Ian~P.~Roberts, and Sriram Vishwanath\\GenXComm, Inc.\\ hardik@genxcomm.com}

\maketitle

\begin{abstract}
% Radar and communication systems have recently been used extensively at \mmwave frequencies for their wide bandwidths. While there exist challenges associated with operating at such high frequencies, both radar and communication systems are offered gains unavailable at lower frequencies, especially in the crowded sub-6 GHz regime. Without proper coordination, conventional radar and communication devices cannot coexist in the same bands due to the incurred interference between the two. Conventionally, simultaneous operation is performed by either frequency duplex resulting in lower throughput and resolution or time duplex resulting in significant increase in latency. In this paper, we present a solution to actively cancel interference between mmWave radar and communication system allowing simultaneous operation in the same frequency. 
%strict allocation of frequency bands prevents such interference but is an ineffective use of the available spectrum. Furthermore, allocation of 60 GHz bands for \ism applications will inevitably introduce interference between devices attempting to take advantage of its wide bandwidths.

The wide bandwidths available at \mmwave frequencies have offered exciting potential to wireless communication systems and radar alike. Communication systems can offer higher rates and support more users with \mmwave bands while radar systems can benefit from higher resolution captures. This leads to the possibility that portions of \mmwave spectrum will be occupied by both communication and radar (e.g., 60 GHz \gls{ism} band). 
This potential coexistence motivates the work of this paper, in which we present a design that can enable simultaneous, in-band operation of a communication system and radar system across the same \mmwave frequencies. To enable such a feat, we mitigate the interference that would otherwise be incurred by leveraging the numerous antennas offered in \mmwave communication systems. Dense antenna arrays allow us to avoid interference spatially, even with the hybrid beamforming constraints often imposed by \mmwave communication systems. Simulation shows that our design sufficiently enables simultaneous, in-band coexistence of a \mmwave radar and communication system.

% Consumer and military applications have the potential of operating colocated \mmwave radios and \mmwave radars, 

\end{abstract}

% \fi
\IEEEpeerreviewmaketitle

\glsresetall

\section{Introduction} \label{sec:introduction}
Future wireless networks like \fg cellular and IEEE 802.11ad have turned to \mmwave frequencies (e.g., $28$, $37$, $60$ GHz) for next-generation wireless communication. While there exist significant challenges in operating at such high frequencies, the wide bandwidths available at \mmwave are attractive for their potential in offering higher data rates and supporting more users \cite{Andrews_Buzzi_Choi_Hanly_Lozano_Soong_Zhang_2014,Heath_Gonzalez-Prelcic_Rangan_Roh_Sayeed_2016}. The high path loss and directional nature of communication at \mmwave enables densification of the network, providing higher network throughput in populated areas and enabling applications requiring extremely low latency \cite{Andrews_Zhang_Durgin_Gupta_2016}.
In addition to its bandwidth and propagation characteristics, \mmwave communication necessitates the use of dense antenna arrays to achieve sufficient link margin due to the severe path loss and poor diffraction. % Conveniently, since antenna size decays with increases in frequency, dense antenna arrays at \mmwave are manageable.

In addition to communication systems, radar has also begun taking advantage of the wide bandwidths offered at \mmwave frequencies. Automotive radar ($77$ GHz) and consumer applications ($24$ and $60$ GHz) have introduced radar to many new areas beyond its ubiquitous use in defense, weather monitoring, and remote sensing.

As communication and radar systems begin to leverage \mmwave frequencies, it is only with proper coordination or creative solutions that these systems can avoid interfering with one another. Strict spectrum allocation certainly has its place for orthogonalizing applications in the frequency domain. For example, \fg bands have been allocated by the \gls{fcc} at $28$ and $37$ GHz while automotive radar operates in its own $77$ GHz band. Of course, the coexistence of these two is not of concern thanks to strict spectrum allocation.

We, however, consider the case when a \mmwave radio and a \mmwave radar attempt to operate within a single band (i.e., over the same frequencies). For instance, consider the $60$ GHz \gls{ism} band: a \mmwave radio (e.g., IEEE 802.11ad) and a \mmwave radar (e.g., \cite{ti6843}). If these two devices operate in each other's presence, the interference incurred may be otherwise prohibitive, leaving one or both devices virtually inoperable. 

While coordination via some flavor of \gls{fdd} may be a possible route to avoid this interference, restricting the bandwidth of the radar and communication system would defeat the point of having operated at \mmwave for its wide bandwidths. Proper coordination between a radar and communication system may enable \gls{tdd} to avoid interference, though this would introduce latency at each device and would be quite difficult to implement practically. 

In this paper, we propose a design to enable simultaneous operation of a \mmwave radar and a \mmwave communication system. This means that radar operation and radio communication will be able to operate at the same time and over the same frequencies while in the presence of one another. In fact, our design is specifically for the case when a radio and radar are colocated. Rather than using time or frequency to orthogonalize the devices, we instead choose to separate in space. Leveraging \mimo communication techniques will allow the radar to operate free from interference during a radio's transmission. Likewise, during reception by the radio, the radar's interference will be mitigated. We present our design along with simulation results that indicates that our work has promise in enabling in-band coexistence of \mmwave communication and radar.

% What would otherwise likely be highly detrimental, 

% The joint radar and communication systems are finding applications beyond defense environment.

% This interference mitigation is particularly applicable to 60 GHz ISM with IEEE 802.11ad and consumer radar applications. Also applicable to 77 GHz automotive radar applications (not sure if they have comms here though).

% We use the following notation. 
\textit{Notation:}
We use bold uppercase, \mat{A}, to represent matrices and bold lowercase, \vec{a}, to represent column vectors. 
We use $(\cdot)$\ctrans, \frobenius{\cdot}, and $\ev{\cdot}$ to represent conjugate transpose, Frobenius norm, and expectation, respectively. 
We use $\left[\mat{A}\right]_{i,j}$ to denote the element in the $i$th row and $j$th column of \mat{A}. 
We use $\left[\mat{A}\right]_{i,:}$ and $\left[\mat{A}\right]_{:,j}$ to denote the $i$th row and $j$th column of \mat{A}. 
We use $\cgauss{\vec{m}}{\mat{R}}$ as a multivariate circularly symmetric complex Normal distribution with mean $\vec{m}$ and covariance $\mat{R}$.
% We use $(\cdot)\todB$ to represent a quantity in unit of \gls{db}. 
% Finally, we use \eyevec{N} to denote an $N$-element column vector of ones and \eyemat{N} to denote an $N \times N$ identity matrix.

\begin{figure*}[!t]
\centering
\subfloat[Transmitting from $i$ to $j$.]{\includegraphics[width=0.4\linewidth]{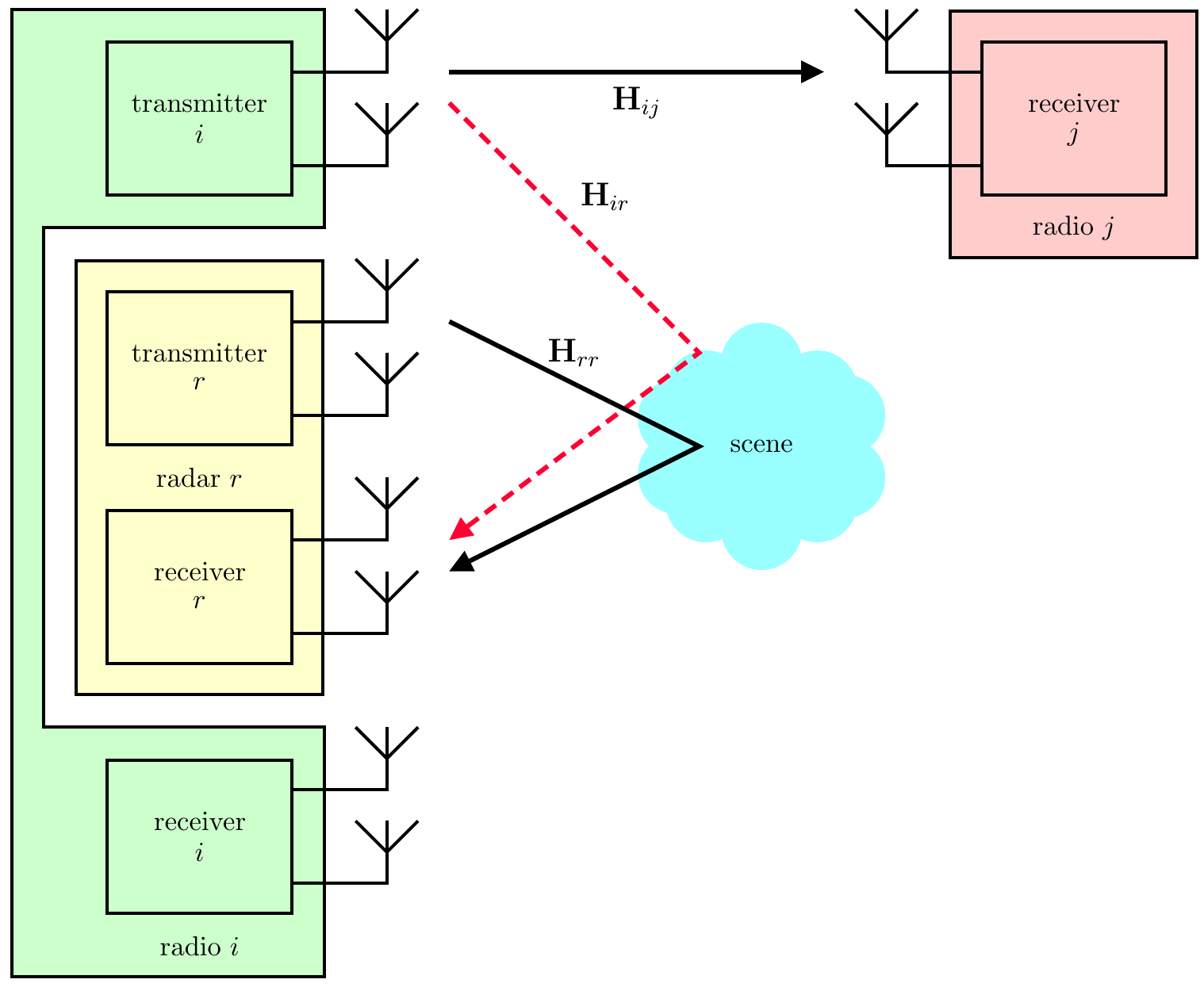}
\label{fig:system-1}}
\hfil
\subfloat[Receiving at $i$ from $k$.]{\includegraphics[width=0.4\linewidth]{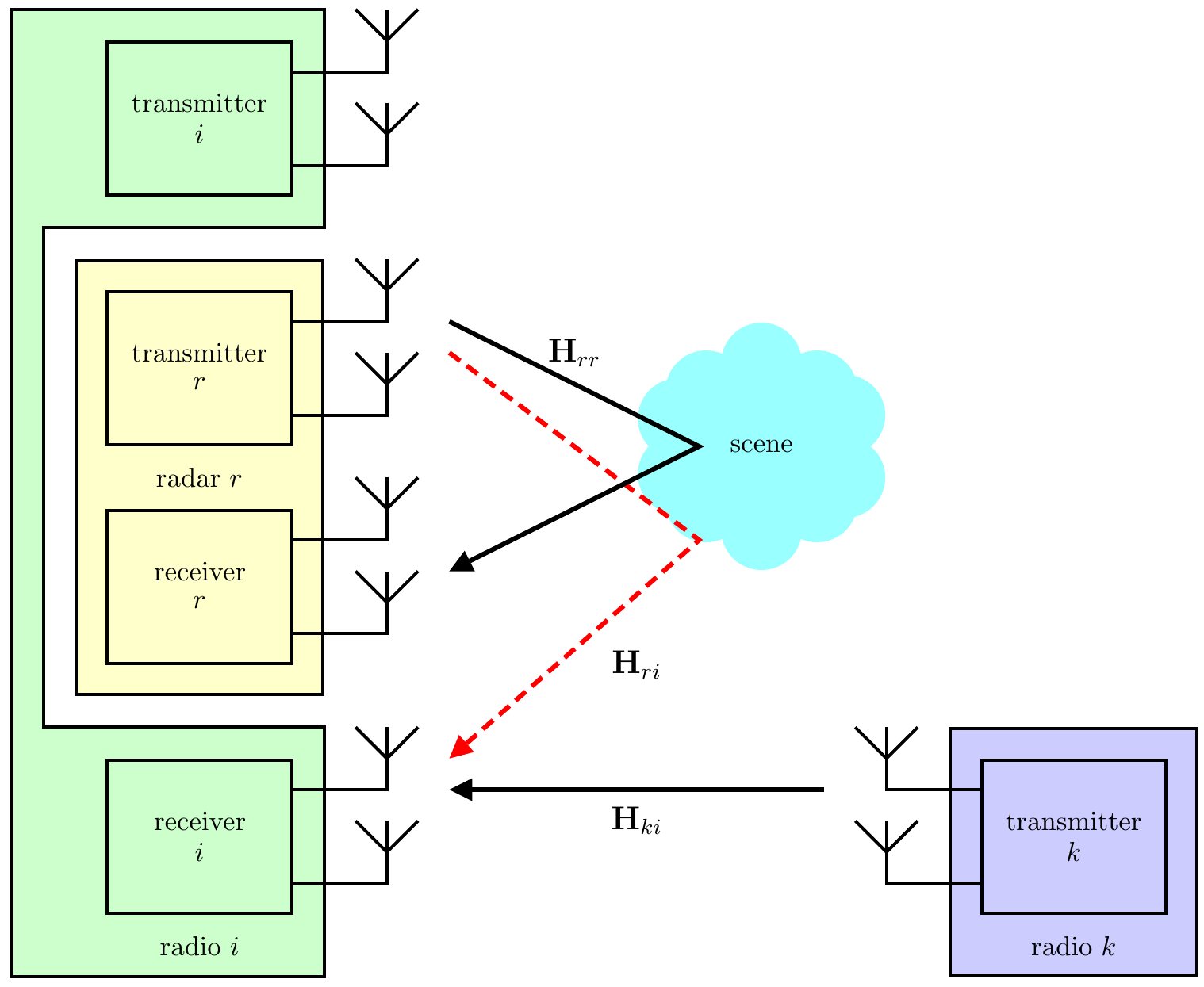}
\label{fig:system-2}}
\caption{Colocation of a \mmwave radar and \mmwave radio introduces interference when operating simultaneously and in-band.}
\label{fig:system}
\end{figure*}

\section{System Model} \label{sec:system-model}

% \begin{figure}[t]
% 	\centering
% 	\includegraphics[width=\linewidth]{fig/system-fig-hj/system-a.png}
% 	\caption{Colocation of a \mmwave radar and \mmwave radio introduces interference when operating simultaneously in the same frequency band.}
% 	\label{fig:system-1}
% \end{figure}

% \begin{figure}[t]
% 	\centering
% 	\includegraphics[width=\linewidth]{fig/system-fig-hj/system-b.png}
% 	\caption{Colocation of a \mmwave radar and \mmwave radio introduces interference when operating simultaneously in the same frequency band.}
% 	\label{fig:system-2}
% \end{figure}

% \begin{figure}[t]
% 	\centering
% 	\includegraphics[width=\linewidth]{fig/system-tx}
% 	\caption{Colocation of a \mmwave radar and \mmwave radio introduces interference at the radar's receiver during transmit mode.}
% 	\label{fig:system-1}
% \end{figure}

% \begin{figure}[t]
% 	\centering
% 	\includegraphics[width=\linewidth]{fig/system-rx}
% 	\caption{Colocation of a \mmwave radar and \mmwave radio introduces interference at the radio's receiver during receive mode.}
% 	\label{fig:system-2}
% \end{figure}

We consider the systems shown in \figref{fig:system-1} and \figref{fig:system-2}.
In the scenario depicted in \figref{fig:system-1}, a \mmwave radar $r$ is observing the scene while a \mmwave radio $i$ transmits to another \mmwave radio $j$. We assume radar $r$ and radio $i$ are colocated and that radio $j$ is relatively distanced from the two. By design, we assume transmissions made by the radar and by radio $i$ are simultaneous and over the same frequencies. As $i$ transmits to $j$, a portion of its transmit signal couples into the receiver of the radar. 

In the scenario depicted in \figref{fig:system-2}, a \mmwave radar $r$ is observing the scene while a \mmwave radio $i$ receives from another \mmwave radio $j$. We again assume radar $r$ and radio $i$ are colocated and that radio $k$ is relatively distanced from the two. Again, by design, we assume transmissions made by the radar and by radio $k$ are simultaneous and over the same frequencies. As $i$ receives from $k$, a portion of the radar's transmit signal couples into the receiver of $i$. 

Our proposed design in Section~\ref{sec:proposed-design} seeks to mitigate the interference incurred in both scenarios to enable simultaneous in-band operation of the \mmwave radio $i$ and the \mmwave radar $r$. We remark that radios $j$ and $k$ can be separate devices or can comprise a single device with transmit/receive capability (i.e., $j$ and $k$ are the same device), though we consider the general case that they are separate devices.

\begin{figure*}[!t]
	% ensure that we have normalsize text
	\normalsize
	% Store the current equation number.
	\centering
	\includegraphics[width=0.85\linewidth]{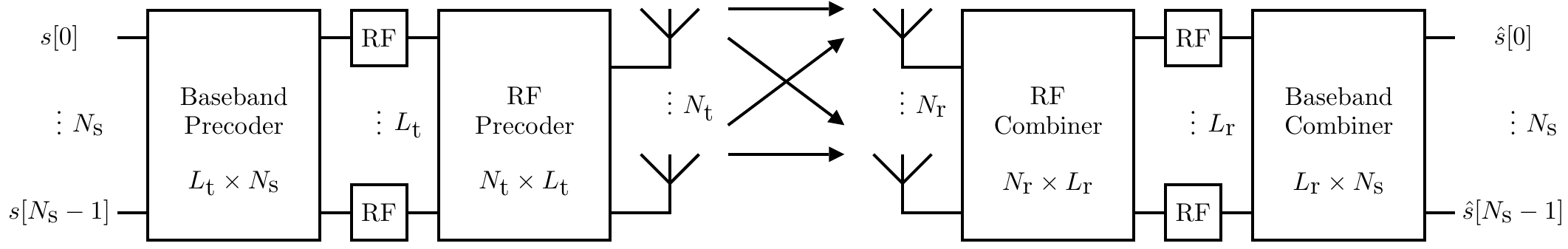}
	\caption{Hybrid beamforming structure commonly used in \mmwave \mimo communication systems.}
	\label{fig:radio}
	% IEEE uses as a separator
	% \hrulefill
	% The spacer can be tweaked to stop underfull vboxes.
	\vspace*{4pt}
\end{figure*}

We assume that the radar and radios are all operating over the same band of frequencies (i.e., in-band). We assume the two scenarios depicted in \figref{fig:system-1} and \figref{fig:system-2} are duplexed using \gls{tdd}. During a given scenario, however, radar and radio operation are simultaneous (i.e., using the same time-frequency resource).

% As an example, consider the case when a consumer device is operating a radar and radio in the $60$ GHz \gls{ism} band. Perhaps the radar is the \ti IWR6843 \gls{fmcw} radar \cite{ti6843} whereas the radio operates using the IEEE 802.11ad standard---both operating simultaneously over the same frequencies. We simply use this example for illustration; our design does not rely on a specific choice of \mmwave carrier frequency or bandwidth.

In traditional communication and radar systems, it is often impossible to coexist in the fashion we have described due to the incurred interference.
The received interference at the radar and at radio $i$ would likely be prohibitively strong, degrading the fidelity of a desired receive signal and potentially making successful reception virtually impossible. 
A common solution to avoid this interference is to impose strict spectrum allocation of radar operation and \mmwave communication, separating the two in the frequency domain. Radar and communication could also be duplexed in time, though this is much more difficult to implement. 
While strict spectrum allocation is certainly a solution to mitigate interference, it comes at the cost of spectrum usage (i.e., bandwidth).
We instead suggest that, with our design, radio and radar operation can operate simultaneously while sharing the same band. If successful, such a scheme would certainly be more favorable, especially in crowded regimes of spectrum.

% However, we suggest that transmission and reception exhibited in \figref{fig:system} can be achieved on the same time-frequency resource with our proposed design. This would offer significant benefits to the communication system, allowing it to operate in a \fd fashion, potentially doubling the spectral efficiency, all while also enabling a colocated radar to continuously observe the scene. Communication and radar operation can be continuously operate without interference with our design meaning strategies such as \gls{tdd}, \gls{fdd}, or strict spectrum allocation are no longer necessary for radar-communication coexistence.

% We make the important distinction that our design considers a communication system using \gls{tdd}, but the radar does not follow this duplexing---it can continuously operate unaffected by the radios' operation.

\subsection{Modeling the Radar}
We assume the \mmwave radar has an array of $\Nt\node{r}$ transmit antennas and $\Nr\node{r}$ receive antennas. We acknowledge that \mimo radar techniques can be used to take advantage of transmit and receive diversity and that it may be the case that only one transmit antenna is active at a time via duplexing transmission from each antenna sequentially in time (i.e., antenna selection). We remark that our model and design comply with such a technique with the appropriate considerations.
We further acknowledge that transmit and/or receive beamforming may be used by the radar. Our design is not reliant on specific (if any) beamforming used by the radar. In fact, our design could potentially be enhanced if we assume the radar performs transmit and/or receive beamforming.

% Another paper could be in designing spatial estimation for radar in the presence of \mmwave radio. (build the angular estimation vectors (beamformers) so that they reject the interference channel).

% We assume that transmit beamforming is possible by use of an analog (i.e., \rf) beamformer that uses phase shifters but lacks amplitude control. We further remark that these phase shifters may be subject to phase quantization, though our design is not concerned with such.

\subsection{Modeling the Radar Channel}
The scene that the radar is observing we term the ``radar channel''.
We model the $\Nr\node{r} \times \Nt\node{r}$ radar channel $\channel_{rr}$ as the combination of reflections from $\Np$ point targets in the scene based on the model in \eqref{eq:channel-radar} \cite{preeti}. Associated with the $p$th point target reflection is the small-scale gain $\beta_{p}$ capturing the target's \gls{rcs}, the round-trip delay of the reflection $\tau_p$, the \gls{aod} $\aod_{p}$ from the radar's transmit array to the target, and the \gls{aoa} $\aoa_{p}$ from the point target to the radar's receive array. The outer product $\rxresponsevectorradar(\aoa_{p}) \txresponsevectorradar\ctrans(\aod_{p})$ captures the array response of the radar for the \gls{aod} and \gls{aoa}.
\begin{align}
\channel_{\mathrm{rad}} = \sum_{p=1}^{\Np} \beta_{p} \exp(-\j 2\pi f_c \tau_p) \rxresponsevectorradar(\aoa_{p}) \txresponsevectorradar\ctrans(\aod_{p}) \label{eq:channel-radar}
\end{align}
We remark that while the scene is likely comprised of continuous/smooth reflectors, we can discretize the cumulative reflection into the combination of numerous point reflections. The large-scale gain of the radar channel is captured in an \gls{snr} term that we will formalize shortly.

\subsection{Modeling the Radios}
As is common in \mmwave \mimo communication, we assume all radios employ hybrid analog/digital beamforming where \mimo precoding and combining are each accomplished using a baseband stage and a \rf stage as shown in \figref{fig:radio}. Specifically, we assume fully-connected hybrid beamforming is used and that the \rf beamformer has phase control but lacks amplitude control \cite{Heath_Gonzalez-Prelcic_Rangan_Roh_Sayeed_2016}.

In the following definitions, let $m\in\{i,j,k\}$ represent the index of a radio in our system.
Let $\Nt\node{m}$ $(\Nr\node{m})$ be the number of transmit (receive) antennas at radio $m$.
Let $\Nrftx\node{m}$ $(\Nrfrx\node{m})$ be the number of transmit (receive) \rf chains at radio $m$.

Let $\prebb\node{m}$ and $\prerf\node{m}$ be the baseband precoding matrix and \rf precoding matrix used for transmission by radio $m$.
Let $\combb\node{m}$ and $\comrf\node{m}$ be the baseband combining matrix and \rf combining matrix used for reception by radio $m$.
To save on cost, power, and complexity, we assume (as is common) that the entries of $\prerf\node{m}$ and $\comrf\node{m}$ are required to have unit magnitude, capturing phase control but lack of amplitude control.

Let $\symbvec\node{m}$ be the $\Ns \times 1$ symbol vectors intended for radio $m$, where we have assumed $\Ns$ symbol streams are being sent on both communication links (from $i$ to $j$ and from $k$ to $i$). Let $\symbvec\node{m}$ have zero mean and $\ev{\symbvec\node{m}\symbvec\ctransnode{m}} = \frac{1}{\Ns}\mat{I}$.

We impose the following uniform power allocation across streams. To do this, we normalize the baseband precoder for each stream such that 
\begin{align}
\frobeniustwo{\prerf\node{m}\entry{\prebb\node{m}}{:,\ell}} = 1 \ \forall \ \ell \in [0,\Ns-1],
\end{align}
which ensures that
\begin{align}
\frobenius{\prerf\node{m}{\prebb\node{m}}}^2 = \Ns.
\end{align}

\subsection{Modeling the Communication Channels}
Let $\channel_{ij}$ be the $\Nr\node{j} \times \Nt\node{i}$ channel matrix from radio $i$ to radio $j$. 
Let $\channel_{ki}$ be the $\Nr\node{i} \times \Nt\node{k}$ channel matrix from radio $k$ to radio $i$. We employ the ray/cluster (Saleh-Valenzuela) \mmwave channel representation shown in \eqref{eq:clustered-channel-model} to model both of these channels.
In this model, \mmwave propagation is captured as a sum of discrete rays.
An $\Nr\times\Nt$ channel is a sum of the contributions from \numclust scattering clusters, each of which contributes \numrays propagation paths \cite{Mendez-Rial_Rusu_Gonzalez-Prelcic_Alkhateeb_Heath_2016}. 
\begin{equation} \label{eq:clustered-channel-model}
\mat{H} = \sqrt{\frac{\Nt \Nr}{\numrays \numclust}} \sum_{u=1}^{\numclust} \sum_{v=1}^{\numrays} \beta_{u,v} \rxresponsevector(\aoa_{u,v}) \txresponsevector\ctrans(\aod_{u,v})
\end{equation}
In \eqref{eq:clustered-channel-model}, $\rxresponsevector(\aoa_{u,v})$ and $\txresponsevector(\aod_{u,v})$ are the antenna array responses 
at the receiving radio and transmitting radio, respectively, for ray $v$ within cluster $u$ which has some \gls{aoa}, $\aoa_{u,v}$, and \gls{aod}, $\aod_{u,v}$. Each ray has gain $\beta_{u,v} \sim \cgauss{0}{1}$. The normalization outside the summations ensures that $\ev{ \frobeniustwo{ \mat{H} }} = \Nt\Nr$. 
% Note that we have considered a single angular dimension (e.g., azimuth) for the \glspl{aoa} and \glspl{aod} for simplicity and to agree with our use of \glspl{ula} in simulation.

\subsection{Modeling the Interference Channels}
As shown in \figref{fig:system-1} and \figref{fig:system-2}, there are two interference channels: one when transmitting from $i$ and one when receiving from $i$. When radio $i$ is transmitting to radio $j$, a portion of its transmitted energy is reflected back to the radar's receiver by the scene. This received interference is combined with the reflected radar signal, corrupting the radar's observation of the scene, potentially introducing estimation or detection errors. We refer this interference channel from the transmitter at $i$ to the radar as $\channel_{ir} \in \complexpow{\Nr\node{r}}{\Nt\node{i}}$.

Similarly, there exists an interference channel between the radar's transmitter to the receiver of $i$ that corrupts the signal being received from $k$ by $i$. We refer to this channel as $\channel_{ri} \in \complexpow{\Nr\node{i}}{\Nt\node{r}}$.
To model both of these channels we use the radar channel model as shown in \eqref{eq:channel-radar}.

\subsection{MIMO Formulation}
Let $\noisevec\node{m} \sim \cgauss{\mat{0}}{\Rn}$ be a noise vector received by the receive array at radio $m\in\{i,j,k\}$, where we assumed a common noise covariance matrix $\Rn = \sigma^2 \mat{I}$ across nodes. 
% Defining the noise covariance matrix as $\Rn = \sigma^2 \mat{I}$ provides a convenient formulation, where $\sigma^2$ captures the noise variance across.
For $m,n \in \{i,j,k,r\}$, we define the \gls{snr} from $m$ to $n$ as 
\begin{align}
%\snr_{mn} = \frac{\Pt\node{m} \ G^2_{mn}}{\sigma^2_n}
% \snr_{mn} \triangleq \frac{\Pt\node{m} G^2_{mn}}{\sigma^2_n B}, 
\snr_{mn} \triangleq \frac{\Pt\node{m} G^2_{mn}}{\sigma^2} \label{eq:snr},
\end{align}
where $\Pt\node{m}$ is the transmit power amplifier gain at $m$ and $G^2_{mn}$ is the large-scale power gain of the propagation from $m$ to $n$.

With these definitions, we can assemble the following formulations describing the received symbols at the receivers at $j$ and $i$. The received symbol at $j$ is
\begin{align}
\rxvec\node{j} &= \combb\ctransnode{j} \comrf\ctransnode{j} \left(\sqrt{\Pt\node{i}} G_{ij} \channel_{ij} \prerf\node{i} \prebb\node{i} \symbvec\node{j} + \noisevec\node{j} \right) \label{eq:rx-j}
\end{align}
whereas the received symbol at $i$ is
\begin{align}
\rxvec\node{i} = \combb\ctransnode{i} \comrf\ctransnode{i} & \Big( \sqrt{\Pt\node{k}} G_{ki}\channel_{ki} \prerf\node{k} \prebb\node{k} \symbvec\node{i} \nonumber \\ & + \sqrt{\Pt\node{r}} G_{ri} \channel_{ri} \pre\node{r} \symbvec\node{r} + \noisevec\node{i}  \Big) \label{eq:rx-i}.
\end{align}
Note that we have captured the interference at $i$ by the radar $r$ as $\sqrt{\Pt\node{r}} G_{ri} \channel_{ri} \pre\node{r} \symbvec\node{r}$, where $\pre\node{r} \symbvec\node{r}$ represents the radar's effective precoding (e.g., antenna selection or beamforming) and its ``transmitted symbols'' which we model in the same way as $\symbvec\node{m}$. More explicitly, only the portion of the radar's transmit signal that corrupts the \textit{sampled} receive signal at the receiver of $i$ are of concern, and thus, we abstract out the actual radar's waveform. In fact, our design is completely agnostic to the actual radar waveform---we merely rely on knowledge of the radar's estimate of the scene.

\subsection{Remarks}
All channels are \mimo channels, meaning they are captured as matrices. Furthermore, we assume they are all frequency-flat. We remark that we can accommodate frequency-selectivity by designing on a per subcarrier basis similar to that in \cite{fsbfc}. Considering frequency-flat channels will simplify our exposition. As indicated by \eqref{eq:rx-j} and \eqref{eq:rx-i}, we do not consider inter-user interference between radios $i$, $j$, and $k$---supported by high path loss and directivity at \mmwave.

\section{Proposed Design} \label{sec:proposed-design}

\begin{figure*}[!t]
	% ensure that we have normalsize text
	\normalsize
	% Store the current equation number.
	\setcounter{mytempeqncnt}{\value{equation}}
	% Set the equation number to one less than the one
	% desired for the first equation here.
	% The value here will have to changed if equations
	% are added or removed prior to the place these
	% equations are referenced in the main text.
	\setcounter{equation}{12}
	\begin{gather}
	\prebb\node{i} = \entry{\left(\channeleff\ctrans_{ij} \combb\node{j} \combb\ctransnode{j} \channeleff_{ij} + \frac{\snr_{ir}}{\snr_{ij}} \channeleff\ctrans_{ir} \channeleff_{ir} + \frac{\Ns}{\snr_{ij}}\mathbf{I} \right)\inv \channeleff\ctrans_{ij} \combb\node{j}}{:,0:\Ns-1}  \label{eq:rzf} \\ 	\setcounter{equation}{16}
	\combb\node{i} = \entry{\left( \channeleff_{ki} \prebb\node{k} \prebb\ctransnode{k} \channeleff\ctrans_{ki} + \frac{\snr_{ri}}{\snr_{ki}} \channeleff_{ri} \channeleff\ctrans_{ri} + \frac{\Ns}{\snr_{ki}}\mathbf{I} \right)\inv \channeleff_{ki}\prebb\node{k}}{:,0:\Ns-1}  \label{eq:lmmse}
	\end{gather}
	% Restore the current equation number.
	\setcounter{equation}{15}
	\setcounter{equation}{\value{mytempeqncnt}}
	% IEEE uses as a separator
	\hrulefill
	% The spacer can be tweaked to stop underfull vboxes.
	\vspace*{4pt}
\end{figure*}

We now present a design that seeks to mitigate the interference introduced by the channels $\channel_{ir}$ and $\channel_{ri}$. The first stage of our design seeks to mitigate the interference imposed by the transmitter at $i$ onto the radar via $\channel_{ir}$. The second stage of our design seeks to mitigate the interference imposed by the radar's transmitter onto the receiver at $i$ via $\channel_{ri}$. To mitigate these sources of interference, we leverage the antenna arrays used at the transmitter and receiver of $i$. % In our design, we assume that perfect \gls{csit} and \gls{csir} is present at each radio. 

% In our design, we make the following assumptions. We assume the number of transmit and receive \rf chains at $i$ is at least BLANK. 

\subsection{Interference Channel Knowledge}
We make the important assumption in our design that radar $r$ and radio $i$ are colocated and are cooperative to the following extent. Being colocated---more specifically, having the two transmitters of the radar and radio $i$ colocated and the two receivers of the radar and radio $i$ colocated---allows us to make the following assumptions.
% \begin{align}
% \channel_{ir} \approx \channel_{rr} \approx \channel_{ri} \label{eq:int-approx-radar}
% \end{align}
It is the goal of the radar to observe the scene, effectively estimating $\channel_{rr}$. Using the observation of the scene, \gls{aod} and \gls{aoa} can be estimated using \mimo radar principles. This \gls{aod} and \gls{aoa} information, along with the estimated gain along those directions, allows us to estimate the interference channels. Knowing the array response of the radio $i$ and of the radar's receiver, the channels $\channel_{ir}$ and $\channel_{ri}$ can be synthesized as according to \eqref{eq:clustered-channel-model}.
For our design, we assume perfect estimation of $\channel_{ir}$ and $\channel_{ri}$ using this method which is passed to radio $i$, allowing us to design on full \gls{csi} of the interference channels. We also assume $\snr_{ir} \propto \snr_{ri} \approx \snr_{rr}$, which are all three roughly equivalent up to a scaling based on the transmit power disparity between the radar and $i$ as indicated by \eqref{eq:snr}.

\subsection{Beamtraining Phase}
In practical \mmwave communication systems, there are significant challenges associated with establishing a link between two devices given the path loss faced at \mmwave frequencies. While beamforming transmission and reception with dense antenna arrays can provide sufficient link margin, the steering direction is initially unknown to both parties. This has introduced the concept of beamtraining \cite{Heath_Gonzalez-Prelcic_Rangan_Roh_Sayeed_2016} where establishing or maintaining a link between two \mmwave radios is done so via a beamspace search. In this search, the two radios of a given link perform a sweep through space measuring the received power for different pairs of beamformers. After sufficient measurements have been made, the parties agree on a pair of \rf beamformers that offer sufficient link margin for communication, upon which further precoding and combining can be done in baseband using the effective channel as seen through the \rf beamformers.

In our design, we assume beamtraining has been performed, though we don't rely on a particular beamtraining strategy. Having undergone beamtraining, the \rf beamformers are set at all radios. We fix $\prerf\node{i}$, $\comrf\node{j}$, $\prerf\node{k}$, and $\comrf\node{i}$ to those determined during beamtraining.
Having fixed the \rf beamformers at a all radios, we assume perfect channel estimation can be done on the relatively small channels seen by the baseband beamformers. Channels that were once of dimension $\Nr \times \Nt$, for example, are reduced to $\Nrfrx \times \Nrftx$---a much smaller value at \mmwave where the number of antennas is large, but the number of \rf chains is small. These reduced communication channels are now
\begin{align}
    \channeleff_{ij} &\triangleq \comrf\ctransnode{j} \channel_{ij} \prerf\node{i}  \in \complexpow{\Nrfrx\node{j}}{\Nrftx\node{i}} \\
    \channeleff_{ki} &\triangleq \comrf\ctransnode{i} \channel_{ki} \prerf\node{k} \in \complexpow{\Nrfrx\node{i}}{\Nrftx\node{k}}
\end{align}
which we assume are fully known at both ends of their respective links.

\subsection{Mitigating Interference onto the Radar}
Having performed beamtraining, we now consider the case shown in \figref{fig:system-1} when radio $i$ is transmitting while radar $r$ is observing the scene. In other words, we consider the \gls{tdd} time slot corresponding to transmission from $i$ to $j$.
We set the baseband combiner at $j$ as the left singular vectors corresponding to the $\Ns$ strongest singular values upon taking the \gls{svd}
\begin{align}
    \channeleff_{ij} = \mat{U}_{ij} \mat{\Sigma}_{ij} \mat{V}_{ij} \label{eq:svd-ij}
\end{align}
where $\mat{\Sigma}_{ij}$ has decreasing singular values along its diagonal. Explicitly, we assign the combiner as
\begin{align}
    \combb\node{j} = \entry{\mat{U}_{ij}}{:,0:\Ns-1}.
\end{align}

The channel $\channel_{ir}$ from the transmitter of $i$ to the radar's receiver introduces undesired interference, potentially corrupting the radar's observations of the scene. Similarly, having fixed the \rf beamformer at the transmitter of $i$, we can consider the effective interference channel
\begin{align}
    \channeleff_{ir} = \channel_{ir} \prerf\node{i} \in \complexpow{\Nr\node{r}}{\Nrftx\node{i}}
\end{align}
which can be computed since we have knowledge of $\channel_{ir}$ and of our choice of $\prerf\node{i}$.

We now seek to design the baseband precoder $\prebb\node{i}$ to transmit into a desired channel $\channeleff_{ij}$ while avoiding pushing interference onto $\channeleff_{ir}$. This naturally motivates a \gls{lmmse} solution, which we write as \eqref{eq:rzf}, commonly referred to as a \gls{rzf} transmitter. Upon normalizing our precoders according to our power constraint, our design for this scenario is complete.

\addtocounter{equation}{1}

\subsection{Mitigating Interference onto the Receiver at $i$}

We now consider the case shown in \figref{fig:system-2} when radio $i$ is receiving while radar $r$ is observing the scene. In other words, we consider the \gls{tdd} time slot corresponding to transmission from $k$ to $i$.
We set the baseband precoder at $k$ as the right singular vectors corresponding to the $\Ns$ strongest singular values upon taking the \gls{svd}
\begin{align}
    \channeleff_{ki} = \mat{U}_{ki} \mat{\Sigma}_{ki} \mat{V}_{ki} \label{eq:svd-ki}
\end{align}
where $\mat{\Sigma}_{ki}$ has decreasing singular values along its diagonal. Explicitly, we assign the precoder as
\begin{align}
    \prebb\node{k} = \entry{\mat{V}_{ki}}{:,0:\Ns-1}.
\end{align}

The channel $\channel_{ri}$ from the transmitter of the radar to the receiver of $i$ introduces undesired interference to, potentially corrupting the radio's reception from $k$. Having fixed the \rf beamformer at the receiver of $i$ following beamtraining, we can consider the effective interference channel
\begin{align}
    \channeleff_{ri} = \comrf\ctransnode{i} \channel_{ri} \in \complexpow{\Nrfrx\node{i}}{\Nt\node{r}}
\end{align}
which can be computed since we have knowledge of $\channel_{ri}$ and of our choice of $\comrf\node{i}$.

We now seek to design the baseband combiner $\combb\node{i}$ to receive from the desired channel $\channeleff_{ki}$ while avoiding receiving interference from $\channeleff_{ri}$. This again motivates a \gls{lmmse} solution, which we write as \eqref{eq:lmmse}. Upon normalizing our precoders according to our power constraint, our design for this scenario is complete.

\addtocounter{equation}{1}

\subsection{Remarks}
We would like to point out that our design is relatively agnostic of the type of radar being used; only operation of radio $i$ is altered with our design. We do make the important point that our design hinges on having sufficient dimensions in the effective channels for avoiding interference. For this reason, increasing the number of \rf chains at the transmitter or receiver of $i$ will improve interference mitigation. Alternatively, a reduction in the number of transmit and receive antennas at the radar would also improve our design's ability to avoid interference. To completely mitigate interference during transmission of $\Ns$ data streams, we require at least $\Nrftx\node{i} \geq \Nr\node{r} + \Ns$.
Similarly, To completely mitigate interference during reception of $\Ns$ data streams, we require at least $\Nrfrx\node{i} \geq \Nt\node{r} + \Ns$.

% Our design could be extended to enable full-duplex operation at radio $i$ allowing it to transmit and receive simultaneously while the radar also operates. Inspiration for such strategies can be found in \cite{ipr-bfc-2019-globecom}.

\section{Simulation and Results} \label{sec:simulation-results}

To evaluate our design we simulated the scenario using the following parameters in a Monte Carlo simulation. We use \glspl{ula} for all arrays where the number of transmit antennas and receive antennas is $32$ at all radios. We use $3$ transmit antennas and $4$ receive antennas at the radar, based on the \ti IWR6843 $60$ GHz radar \cite{ti6843}. We let the number of \rf chains at the receiver of $j$ and the transmitter of $k$ to be $2$. We let the number of \rf chains at the transmitter and receiver of $i$ to be $\Nrftx\node{i} = \Nrfrx\node{i} = 8$. We transmit $\Ns=2$ streams on both communication links. The radar channel $\channel_{rr}$ is populated with $\Np = 600$ point targets uniformly distributed in azimuth and in range (up to $100$ meters). The communication channels are statistically equivalent, each taking a random number of clusters from $[1,6]$ and a random number of rays per cluster $[1,10]$. We let $\snr_{ij} = \snr_{ki}$ for simplicity of interpreting results. We let $\snr_{rr} = 40$ dB (e.g., consider a noise floor of $-80$ dBm and the average reflected power to be $-40$ dBm). During beamtraining for each channel, we take the strongest beam pairs from a \gls{dft} codebook. We assume equal transmit power at the radar and radios.

\subsection{Performance on the Communication Links}
\begin{figure}
    \centering
    \includegraphics[height=6.5cm]{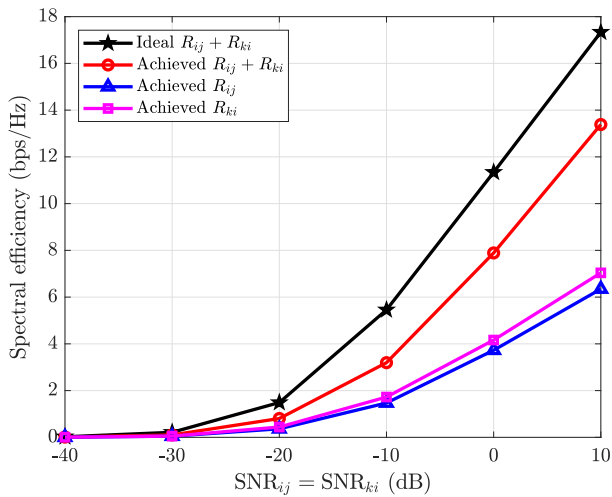}
    \caption{Spectral efficiency as a function of the desired link \glspl{snr}; shown are our achieved link spectral efficiencies and their sum (red) versus the sum spectral efficiency had the radar not been present (black).}
    \label{fig:speceff}
\end{figure}

Our primary metric in evaluating performance on the communication links is spectral efficiency (we omit explicit spectral efficiency expressions due to space constraints). Let $R_{ij}$ and $R_{ki}$ be the spectral efficiencies on the links from $i$ to $j$ and from $k$ to $i$, respectively. Given our system model and assumptions, the optimal transmit strategy ignoring interference is transmitting and receiving along the right and left singular vectors corresponding to the strongest \Ns singular values of the communication channels $\channel_{ij}$ and $\channel_{ki}$. The sum spectral efficiency under such a scheme is shown in \figref{fig:speceff}, which serves as a baseline for evaluating our design. The closer our design approaches this ideal sum spectral efficiency, the better. Our design does quite well considering it mitigates a significant portion of the interference it would otherwise introduce. The spectral efficiency achieved in both links are well balanced as shown in \figref{fig:speceff}. This is expected given the design of the two scenarios are nearly duals of one another; the primary difference is in the number of transmit and receive antennas at the radar.

\subsection{Performance at the Radar}
We use the definition of \gls{sir} in \eqref{eq:sir} as a proxy for performance radar performance. 
\begin{align}
    \sir_{rr} \triangleq {\trace\left\{\channel_{ir} \prerf\node{i} \prebb\node{i} \Rs \prebb\ctransnode{i} \prerf\ctransnode{i} \channel\ctrans_{ir}\right\}}\inv
    \label{eq:sir}
\end{align}
Noise and desired signal power being fixed, $\sir_{rr}$ quantifies solely the interference incurred at the radar due to transmission from radio $i$. Recall that our design tailored $\prebb\node{i}$ to avoid this interference by trading off ideal transmission to $j$ for interference mitigation.

\begin{figure}
    \centering
    \includegraphics[height=6.5cm]{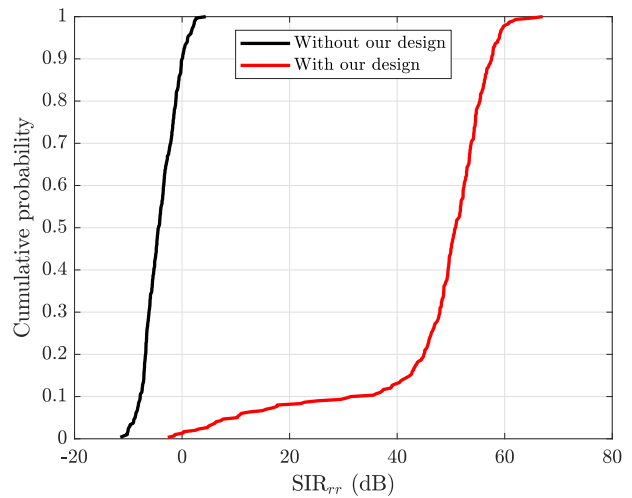}
    \caption{\Gls{cdf} of the \gls{sir} of the radar's desired receive signal when using our design (red) as compared to without (black).}
    \label{fig:cdf}
\end{figure}

The \gls{cdf} of $\sir_{rr}$ can be seen in \figref{fig:cdf}. The \gls{cdf} shown is over 250 iterations from $\snr_{ij} = \snr_{ki} \in \{-40, -30, -20, -10, 0, 10\}$ dB. 
% Desirably, $\sir_{rr}$ would be many orders of magnitude higher than the $\snr_{rr}$ so that the noise would dominate the \gls{sinr}. 
It is clear from \figref{fig:cdf} achieves this, netting $\sir_{rr}$ values that are above $30$ dB over $90$\% of the time.
Keep in mind that this performance is achieved all the while $i$ transmits to $j$ with the spectral efficiency exhibited in \figref{fig:speceff}.
When our design is not used and the presence of the radar is ignored, we can see that the interference is quite strong, as shown in black in \figref{fig:cdf}. In such a case, the interference that is coupled into the radar's receiver is clearly overwhelming. We remark that our definition of \gls{sir} abstracts out the transmit power of $i$ and of the radar, meaning these curves could shift left or right depending on the transmit power disparity between the radar and radio $i$. The gap between the two would shrink when the radar has a higher transmit power than radio $i$ and would widen in the reverse case.

\section{Conclusion} \label{sec:conclusion}

In this paper, we presented a beamforming design that mitigates interference encountered when a colocated \mmwave radio and \mmwave radar operate simultaneously and in-band. Interference that would be otherwise prohibitive is mitigated by \mimo precoding and combining strategies. Simulation indicates that our design sufficiently mitigates interference in transmit and receive scenarios, leaving radar and radio operation unaffected by simultaneous, in-band operation.

% We look forward to extending this work to more sophisticated network scenarios and to formalize methods for radar-assisted interference channel estimation as presented in this work.

% \input{sec-acknowledgement.tex}

% \section*{References} \label{sec:bibliography}
% \printbibliography[heading=none]
\bibliography{refs}

\end{document}